\documentclass[prb,twocolumn,preprintnumbers,amsmath,amssymb,floats,citeautoscript,nobalancelastpage,showpacs]{revtex4-1}

\usepackage{graphicx}
\usepackage{color}
\usepackage{bm}

\begin{document}

\title{High-temperature thermoelectric properties of the double-perovskite ruthenium
oxide (Sr$_{1-x}$La$_x$)$_2$ErRuO$_6$ }

\author{Ryohei Takahashi}
\author{Ryuji Okazaki}
\author{Yukio Yasui}
\altaffiliation[Present address:]{
Department of Physics, Meiji University, 
Kawasaki 214-8571, Japan
}
\author{Ichiro Terasaki}
\email[Email me at:]{
terra@cc.nagoya-u.ac.jp
}

\affiliation{Department of Physics, Nagoya University, Nagoya 464-8602, Japan}

\author{Takaaki Sudayama}
\author{Hironori Nakao}
\author{Yuichi Yamasaki}
\author{Jun Okamoto}
\author{Youichi Murakami}

\affiliation{Condensed Matter Research Center and Photon Factory, 
Institute of Materials Structure Science, 
High Energy Accelerator Research Organization, Tsukuba 305-0801, Japan}

\author{Yoshinori Kitajima}

\affiliation{Photon Factory, Institute of Materials Structure Science, 
High Energy Accelerator Research Organization, 
Tsukuba 305-0801, Japan}

\begin{abstract}
 We have prepared  polycrystalline samples of 
 (Sr$_{1-x}$La$_x$)$_2$ErRuO$_6$
and (Sr$_{1-x}$La$_x$)$_2$YRuO$_6$, 
 and have measured the resistivity, Seebeck coefficient,  
 thermal conductivity, susceptibility and x-ray absorption 
 in order to evaluate the electronic states and thermoelectric properties
 of the doped double-perovskite ruthenates.
 We have observed a large Seebeck coefficient of $-160$ $\mu$V/K
 and a low thermal conductivity of 7 mW/cmK for $x$=0.1 at 800 K in air.
These two values are suitable for efficient  oxide thermoelectrics,
 although the resistivity is still as high as 1 $\Omega$cm.
 From the susceptibility and x-ray absorption measurements, 
 we find that the doped electrons exist as Ru$^{4+}$ in the low spin
 state.
 On the basis of the measured results, the electronic states and the
 conduction mechanism are discussed.
\end{abstract}

\maketitle
\section{Introduction}
Thermoelectrics is a technology that converts heat into electric power
or vice versa through the thermoelectric
phenomena in solids.\cite{mahan1998}
Since this technology is a direct energy conversion in solids, 
it has attracted a renewed interest as a fundamental technology for 
environmentally-friendly energy conversion. 
In particular, thermoelectric power generation has been now considered
as a possible renewable energy resource.

Oxide thermoelectrics has been extensively investigated 
as a promising thermoelectric power generator,
for oxides are stable at high temperatures in air.
Oxides were considered to be poor thermoelectric materials, 
but after the discovery of a large thermoelectric power factor 
in Na$_x$CoO$_2$, some cobalt oxides are recognized as 
good thermoelectric oxides of p-type.\cite{terasaki1997,takahata2000}
In contrast, not yet discovered is an n-type counterpart 
to the cobalt oxides.
Some of the transparent conductors such as ZnO and In$_2$O$_3$
show indeed good thermoelectric performance 
above 1000 K,\cite{ohtaki1996,beradan2008}
but the lattice thermal conductivity is much higher than 
the conventional thermoelectric materials. 
The doped titanates \cite{okuda2001,ohta2005}
and niobates \cite{sakai2010,kobayashi2011,lee2011}
are fairly good  n-type thermoelectric materials at room temperature, 
but they are easily oxidized at high temperature to lose conductivity in
air.

Recently a large Seebeck coefficient and a low thermal conductivity have been
reported in polycrystalline samples of the double perovskite ruthenate
Sr$_2L$RuO$_6$ ($L$;  rare-earth).\cite{aguirre2009}
This particular ruthenate was first synthesized by Donahue and 
McCann,\cite{donohue1977}
whose crystal structure and physical properties were investigated 
by Battle and Wacklyn.\cite{battle1984}
It crystallizes in the $B$-site ordered perovskite 
structure of $A_2BB'$O$_6$,
where the two different cations of 
$L$ and Ru occupy the $B$ and $B'$ sites
to form an NaCl type ordered structure.
As a unique feature, the Ru ion is pentavalent (Ru$^{5+}$)
with the electronic configuration of $(4d)^3$, 
which acts as a local moment of $S=3/2$ to show an antiferromagnetic
order below 26 K in Sr$_2$YRuO$_6$.\cite{cao2001}
When magnetic rare-earth ions occupy the $B$ site,
the transition temperature and the magnetic structure change
depending on the species of the rare-earth 
ions.\cite{doi1999,doi2000,izumiyama2001,rao2006,puche2007}
In addition, a possible high-temperature superconductivity has been discussed in 
Sr$_2L$Ru$_{1-x}$ Cu$_x$O$_6$.\cite{wu1996,blackstead2000,harshman2003}

In this paper, we show the thermoelectric properties  
in polycrystalline samples of Sr$_{2-x}$La$_x$ErRuO$_6$,
in which partial substitution of La for Sr supplies electrons to let
the samples n-type.
The Seebeck coefficient is almost independent of temperature 
above room temperature, whose magnitude
is roughly explained in terms of the Heikes formula.
The thermal conductivity is lower than 10 mW/cmK at 800 K, which is
quite anomalous in comparison with the thermal conductivity 
of other oxides.
X-ray absorption and susceptibility measurements 
have revealed that the Ru$^{4+}$ induced
by La substitution for Sr is in the low spin state, which implies that
the doped electron occupies the upper Hubbard $t_{2g}$ manifolds.
On the basis of the measured data, the electronic states and the conduction
mechanism are discussed.

\section{experimental}
Polycrystalline samples of (Sr$_{1-x}$La$_x$)$_2$ErRuO$_6$ 
($x$ = 0, 0.05, 0.1, 0.2, and 0.3) and
(Sr$_{1-x}$La$_x$)$_2$YRuO$_6$ \ \ ($x$ = 0, 0.1, and 0.2) 
were prepared by solid-state reaction.
Stoichiometric amounts of SrCO$_3$, La$_2$O$_3$, Er$_2$O$_3$, Y$_2$O$_3$, 
and RuO$_2$ were mixed, and the mixture was calcined 
at 900$^{\circ}$C for 12 h in air.
The calcined powder was ground, pressed into a pellet, 
and sintered at 1250$^{\circ}$C for 60 h in air.
X-ray diffraction was measured with a Rigaku RAD-IIC 
(Cu K$\alpha$ radiation), 
and no impurity phases were detected in the prepared samples.

The X-ray absorption spectra were measured at BL-11B KEK-PF, Japan.
All the Ru $L$ edge spectra were measured at room temperature in the
fluorescence yield mode using a photodiode detector.
The base pressure in the chamber was 10$^{-7}$ Torr.
The electrical resistivity $\rho$ was measured using a four-probe method
with a constant current of 1 mA from room temperature to 800 K in air 
with a home-made measurement probe inserted in a cylinder furnace.
The Seebeck coefficient $S$ was measured with a quasi-steady-state method
from room temperature to 800 K in air with a home-made measurement
probe in a cylinder furnace; 
the edges of a bar-shaped sample was pasted to two ceramic plates working as
heat bathes, one of which was heated by a nichrome heater. 
The temperature difference was monitored with a differential
thermocouple made of Pt-PtRh. 
The thermoelectric voltage from the voltage leads was carefully subtracted.
For some of the samples, the Seebeck coefficient was 
measured with a steady-state method using a copper-constantan
differential thermocouple
from room temperature down to 100 K 
in a liquid Helium cryostat. 
The thermal conductivity was evaluated from 
the thermal diffusivity measured 
from room temperature to 800 K in air with a laser flash method
(ULVAC-Riko TC2000)  and the heat capacity measured 
with differential scanning calorimetry (Netzsch DSC404F3) in Ar flow.
The magnetization in field cooling (FC) and zero field cooling (ZFC) 
processes was measured using a superconducting quantum 
interference device magnetometer (Quantum Design MPMS) 
from 5 to 300 K in an applied field of 1 T.

\begin{figure}
\includegraphics[width=8cm,clip]{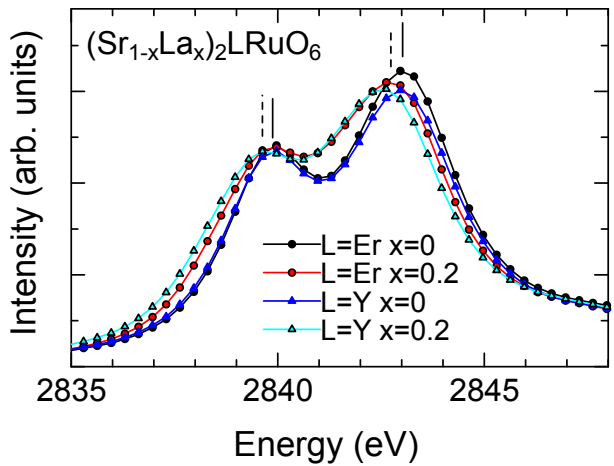}
\caption{
(color online) Ru $L_3$ edge absorption spectra 
of (Sr$_{1-x}$La$_x$)$_2$ErRuO$_6$
and (Sr$_{1-x}$La$_x$)$_2$YRuO$_6$.
Solid and dotted vertical lines denote 
the peak positions for x=0 and 0.2, respectively.
}
\label{fig1}
\end{figure}

\section{results and discussion}
First of all, let us examine the valence state of the Ru ion in the title compound. 
Figure \ref{fig1} shows the Ru $L_3$ edge spectra of 
(Sr$_{1-x}$La$_x$)$_2$ErRuO$_6$
and (Sr$_{1-x}$La$_x$)$_2$YRuO$_6$.
For $x$=0, two peaks are observed around 2840 and 2843 eV,
which evidences the existence of the pentavalent Ru ion as was already
reported.\cite{aguirre2009}
The peak positions and intensities, and accordingly the valence state of the Ru ion, 
are essentially identical between the Er- and Y-based compounds.
For Ru$^{4+}$ oxides such as SrRuO$_3$ and RuO$_2$, 
the $L$ edge spectra show broad peaks at 2838
and 2841 eV \cite{sahu2002}
and by using the $L$ edge spectra we can roughly evaluate the valence state 
from the peak energies.
For the La substituted samples, the peaks are shifted to lower energy 
by around 0.3 eV, indicating that the valence state of the Ru ion 
shifts to a lower valence, which is consistent with 
a naive picture that 
the La substitution for Sr supplies electrons to the system to create 
a tetravalent Ru ion per La.
Comparing the leading edge on the lower-energy side, we find that 
the Ru ions in the $x$=0.2 samples are 
in a similar valence state between Er and Y.
These data thus warrant that the species of the rare-earth ion in the
$B$ site do not affect the valence state of the Ru ions.

\begin{figure}
\includegraphics[width=8cm,clip]{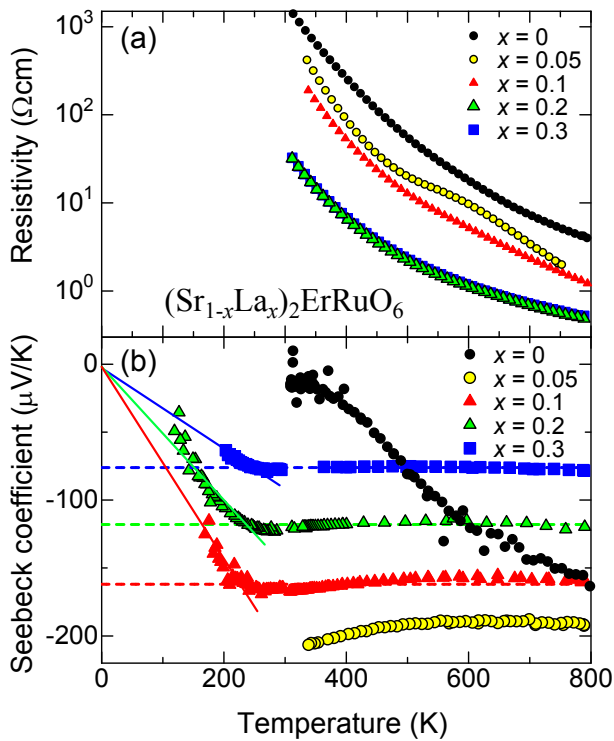}
\caption{
(color online) (a) The resistivity and
 (b) the Seebeck coefficient of
 (Sr$_{1-x}$La$_x$)$_2$ErRuO$_6$ 
($x$ = 0, 0.05, 0.1, 0.2, and 0.3)
 plotted as a function of temperature.
The solid and dotted lines represent 
theoretical  curves (see text).
}
\label{fig2}
\end{figure}

Figure \ref{fig2}(a) shows
the electrical resistivity of (Sr$_{1-x}$La$_x$)$_2$ErRuO$_6$.
All the resistivities decrease with increasing temperature, 
indicating that the samples are nonmetallic.
The magnitude systematically decreases with increasing $x$, 
indicating that the carrier concentration increases with increasing La content.
The resistivity decreases roughly by two orders of magnitude
from $x$=0 to 0.2, but seems saturated near 0.3, 
suggesting the solubility limit of La substitution. 

Figure \ref{fig2}(b) shows 
 the Seebeck coefficient of (Sr$_{1-x}$La$_x$)$_2$ErRuO$_6$.
The sign for all the samples is negative, 
and the magnitude systematically decreases with increasing $x$ except
 for $x$=0.
These results show that the substituted La ion acts as a donor
to supply electrons to the system.
The Seebeck coefficient for $x$=0 is close to zero at 300 K,
and the magnitude increases with increasing temperature,
possibly because small amounts of electrons and holes
inevitably doped through unwanted nonstoichiometry and/or 
impurities show complicated temperature dependence. 
The Seebeck coefficient for $x$=0 is different from that  previously 
reported by Aguirre et al.\cite{aguirre2009}
The magnitude of the Seebeck coefficient of their sample was much larger,
and decreased with increasing  temperature, 
possibly owing to a smaller amount 
of carriers introduced in their sample.
The Seebeck coefficients for $x>0$, on the other hand, 
are essentially independent of temperature above room temperature.

The Seebeck coefficient  below room temperature
decreases with decreasing
temperature, suggesting a $T$-linear behavior, 
although the low-temperature measurements were seriously affected
by the high resistance of the samples.
Since the $T$-linear Seebeck coefficient is a hallmark of metals 
having a finite Fermi energy in the valence band,  
the data clearly indicate 
that the doped samples are essentially metallic 
in the sense that there is a Fermi surface.
Then the nonmetallic resistivity is ascribed to the scattering time, 
which is reasonable because the RuO$_6$ octahedra
are well separated by the ErO$_6$ octahedra to make transfer integrals 
small.\cite{mazin1997}

Now we will evaluate the carrier concentration and the effective mass 
by following the method applied to Nd$_{1.9}$Ce$_{0.1}$PdO$_4$ 
by Shibasaki and Terasaki\cite{shibasaki2006jpsj}.
The $T-$linear Seebeck coefficient can be associated with 
the diffusive term of the Seebeck coefficient 
for a single parabolic band given by 
\begin{equation}
S=-\frac{\pi ^2}{2}\frac{k_B^2T}{eE_F},
\label{equ:SlowT}
\end{equation}
where $E_F$ is the Fermi energy.
From the $T-$linear slope of the measured data, the Fermi energy is
obtained as listed in Table I.

On the other hand, the temperature-independent Seebeck coefficient
can be analyzed with the Heikes formula, an asymptotic expression
of the Seebeck coefficient in the high temperature limit,
where the thermal energy of $k_BT$ is much larger than the band width
or the transfer energy.\cite{chaikin1976}
We examined various forms  for the Heikes formula,
and find that the observed Seebeck coefficient
is well explained by the expression given by 
\begin{equation}
S=-\frac{k_B}{e}\mathrm{ln}\frac{2-p}{p},
\label{equ:ShighT}
\end{equation}
where $p$ is the carrier number per Ru.
From the constant Seebeck coefficient at high temperature, 
the carrier concentration $n$ is evaluated as listed in Table I.
Using $n$ and $E_F$, we further obtain the effective mass $m^*$ 
through the following expression for a single parabolic band  
given by
\begin{equation}
E_F = \frac{\hbar ^2}{2m^*}(3\pi ^2n)^{\frac{2}{3}},
\end{equation}
as listed in Table I.
A crossover temperature around which the Seebeck coefficient 
changes from $T-$linear to temperature-independent can be read off
from Fig. \ref{fig2}(b) to be 200-300 K, which
should correspond to the energy scale for the transfer energy.
Mazin and Singh calculated the band structure of Sr$_2$YRuO$_6$,
and evaluated the transfer energy to be 0.14 eV, which is 
in fact several times larger than the thermal energy of 300 K.
The correlation effects may further reduce the transfer energy, because
the valence bands consist of the lower Hubbard $t_{2g}$ manifolds,
which include the spin-dependent hopping.

The carrier concentration in Table I is roughly proportional to 
the La concentration $x$,
and the magnitude of 10$^{21}$ cm$^{-3}$ is the same
order of the carrier concentration estimated by the assumption 
that one substituted La ion supplies one electron.
This clearly indicates that the substituted La acts as a donor in 
a simplest approximation.
The effective mass is evaluated to be nearly $x$-independent value of 3$m_0$,
which suggests that the electron doping in this system is
rigid-band-like. 
The mobility is formally calculated as $\mu$ = $1/ne\rho\sim 10^{-4}$ 
cm$^2$/Vs at 800 K, which is too small for the Boltzmann
transport where the mean free path must be longer than the lattice spacing.
This is understandable because the doped carriers are well localized in
a RuO$_6$ octahedron, and the electrical conduction occurs via hopping
from one RuO$_6$ octahedron to another.\cite{mazin1997}
The hopping process accompanies a finite activation energy which appears
in the temperature dependence of the resistivity shown in
Fig. \ref{fig2} (a).
Nevertheless we find that  
the mobility is almost independent of the La content, indicating 
that La substitution changes only the carrier concentration
like a rigid-band picture.

\begin{table}
\begin{center}
\caption{Various parameters for (Sr$_{1-x}$La$_x$)$_2$ErRuO$_6$. $E_F$, $n$, and $m^*$ 
are Fermi energy, carrier concentration, and effective mass, respectively.}
\begin{tabular}{c  c c c} \hline\hline
$x$ & $E_F$(meV) & $n$($10^{21}$cm$^{-3}$) & $m^*/m_0$ \\\hline
0.1 & 52.2 & 0.97 & 3.1 \\
0.2 & 71.6 & 1.5 & 3.6 \\
0.3 & 116 & 2.2 & 3.1 \\\hline\hline
\end{tabular}
\label{table:physV}
\end{center}
\end{table}

\begin{figure}
\includegraphics[width=8cm,clip]{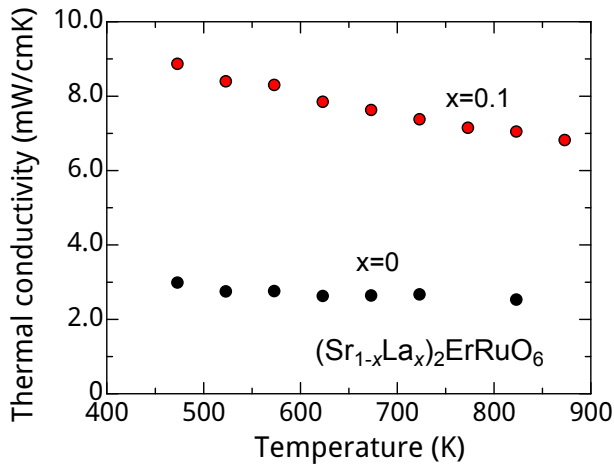}
\caption{
(color online) The temperature dependence of the thermal conductivity 
for (Sr$_{1-x}$La$_x$)$_2$ErRuO$_6$ ($x$ = 0 and 0.1).
}
\label{fig3}
\end{figure}

Figure \ref{fig3} shows the temperature dependence 
of the thermal conductivity of (Sr$_{1-x}$La$_x$)$_2$ErRuO$_6$.
The thermal conductivity slightly decreases with increasing temperature,
with a small magnitude of 3-9 mW/cmK.
The value for $x$=0 is slightly higher than the previously reported
value by Aguirre et al.,\cite{aguirre2009} but their value may come
from porosity of the sample (the sample density of 78-81\%).
In contrast, the density of our samples is larger than 90\%,
and we think that this low thermal conductivity observed here is intrinsic.
The value for $x$=0 is indeed anomalously low, and is close to the minimum thermal
conductivity proposed by Cahill et al.\cite{cahill1992}
Recently Wan et al. \cite{wan2008}
have reported that an oxygen deficient aluminium oxide 
Ba$_2$DyAlO$_5$ exhibits a low value of close to 10 mW/cmK at 1000$^{\circ}$C.
They associated this low value with oxygen deficiency, but 
the present data imply that a different mechanism does exist, for the title compound 
includes no significant oxygen vacancies.
The doped sample of $x$=0.1 is more disordered, and thus the thermal
conductivity is expected to be reduced from $x$=0, 
which is seriously incompatible with the observation.
We note that the electron contribution of the thermal conductivity is evaluated 
to be 0.02 mW/cmK for $x$=0.1 at 800 K using the Wiedemann-Franz law, 
which cannot be a reason of the increase in thermal conductivity 
from $x$=0 to 0.1.
We suggest that 
the double perovskite structure of $A_2BB'$O$_6$ may 
be a key ingredient;
Aguirre et al. \cite{aguirre2009}
found characteristic micro-domain structures in the
transmission electron microscope.
Ohtaki et al. \cite{sugawara2008}
reported that the double perovskite oxide Sr$_2$FeMoO$_6$ 
also shows a low thermal conductivity.
In spite of such low thermal conductivity, 
the dimensionless figure of merit ($ZT$ = $S^2\sigma T/\kappa$) 
of the present ruthenate  remains low ($ZT \sim 10^{-3}$ at 800 K) 
because of the high resistivity.

\begin{figure}
 \includegraphics[width=8cm,clip]{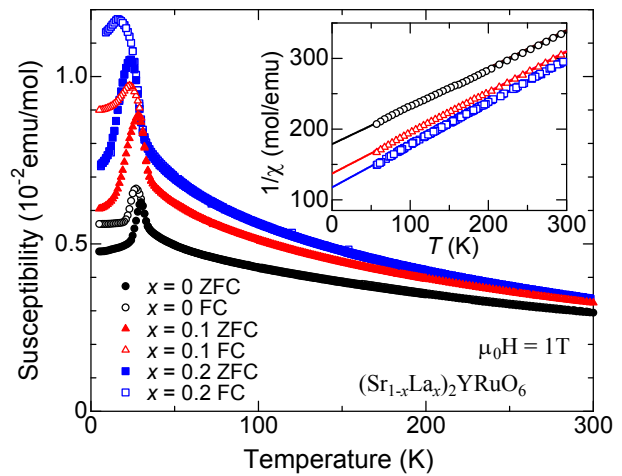}
\caption{
(color online) Zero-field-cooled (ZFC) and field-cooled (FC) 
magnetic susceptibility (defined as $M/H$) as a function 
of temperature at $H$ = 1 T for (Sr$_{1-x}$La$_x$)$_2$YRuO$_6$ ($x$ = 0, 0.1, and 0.2). 
In the inset, the inverse magnetic susceptibility is plotted as a function of temperature. 
The solid line represent Eqs. (\ref{equ:curie}) and (\ref{equ:S}).
}
\label{fig4}
\end{figure}

Let us discuss the electronic properties of the doped electron through the spin
state of the Ru$^{4+}$ ions by measuring the susceptibility.
In order to avoid a large magnetization arising from Er ions,
we used   (Sr$_{1-x}$La$_x$)$_2$YRuO$_6$.
Figure \ref{fig4} shows the temperature dependence of 
magnetic susceptibility for (Sr$_{1-x}$La$_x$)$_2$YRuO$_6$.
The magnetic phase transition is visible at $T_N$ = 34 K for $x$ = 0, 
below which the magnetic susceptibility shows temperature hysteresis,
which is consistent with previous measurements.\cite{battle1984,cao2001}
We analyze the susceptibility from 50 to 300 K 
using the Curie-Weiss law given by
\begin{equation}
\chi = \frac{\mu _{\rm eff}^2{\mu _B}^2}{3k_B(T+\theta )},
\label{equ:curie}
\end{equation}
and evaluate the effective magnetic moment $\mu _{\rm eff}$ 
and the Weiss temperature $\theta$. 
The effective magnetic moment  is 3.87 $\mu_B$/Ru for $x$=0,
which is close to  a theoretical value of $S=3/2$
and also consistent with the previous work\cite{cao2001}.
For $x$ = 0.1 and 0.2, the spin number obtained experimentally 
($S_{\mathrm{exp}}$) is given by an average of Ru$^{5+}$ and Ru$^{4+}$
in a simplest approximation as
\begin{equation}
S_{\mathrm{exp}} = (1-x)S_{\mathrm{Ru}^{5+}} + xS_{\mathrm{Ru}^{4+}},
\label{equ:S}
\end{equation}
where $S_{\mathrm{Ru}^{5+}}$ (=3/2), $S_{\mathrm{Ru}^{4+}}$, 
and $x$ are the spin number of Ru$^{5+}$ , 
the spin number of Ru$^{4+}$, 
and the fraction of Ru$^{4+}$, respectively.
As listed in table \ref{table:sus}, $S_{\mathrm{Ru}^{4+}}$ is 
calculated from $x$ and $S_{\rm exp}$ to be close to unity, 
indicating that Ru$^{4+}$ is in the low-spin state.
Accordingly, we conclude that the conduction band for the doped
electrons in this system is composed of the upper Hubbard $t_{2g}$
manifolds.

We notice that the transition temperature $T_N$ is much smaller 
than the Weiss temperature $\theta$, 
which has been associated with
frustration effects.\cite{mazin1997,kuzmin2003,ravi2008}
We also notice that $\theta$ is anomalously reduced with $x$, 
while $T_N$ does not change much.
Even if some frustration effects may appear in the title compound,
the disorder effects on $T_N$ and $\theta$ are highly difficult to
explain;
The induced Ru$^{4+}$ seems to weaken the spin-spin interaction
($\propto \theta$)
but seems to leave the ordered energy gain ($\propto T_N$) intact.
In addition to the charge transport, the magnetic properties of 
this compound is not trivial, which should be clarified by further investigations.
Singh and Tomy \cite{ravi2008} found two-step magnetic transition (27 and 32 K)
in Sr$_2$YRuO$_6$ from a careful magnetization measurement, 
which implies an existence of two components.  

\begin{table}
\begin{center}
\caption{
Various parameter for (Sr$_{1-x}$La$_x$)$_2$YRuO$_6$. 
$\mu_{\rm eff}$, $\theta$, $S_{\mathrm{exp}}$, 
and $S_{\mathrm{Ru}^{4+}}$ are effective Bohr magnetic moment, 
Curie-Weiss temperature, spin of experimental value, and spin 
of Ru$^{4+}$ calculated Eqs. (\ref{equ:S}).
}
\begin{tabular}{c c c c c} \hline\hline
$x$ & $\mu _{\rm eff}$ ($\mu_B$/Ru) 
& $\theta$ (K) & $S_{\mathrm{exp}}$ & $S_{\mathrm{Ru}^{4+}}$ \\\hline
0 & 3.87 & 338 & 1.5 & - \\
0.1 & 3.73 & 240 & 1.43 & 1.16 \\
0.2 & 3.65 & 196 & 1.39 & 1.23 \\\hline\hline
\end{tabular}
\label{table:sus}
\end{center}
\end{table}

Here we will discuss the electronic states and the conduction mechanism
of the doped Sr$_2$ErRuO$_6$.
Mazin and Singh\cite{mazin1997} calculated the electronic band structure of 
Sr$_2$YRuO$_6$.
Despite the complicated structure, the electronic states can be
quite simply understood; the RuO$_6$ octahedra are responsible
for the valence band and the electrical conduction, which are
isolated from each other by the YO$_6$ octahedra.
Thus the electronic states are essentially understood from the energy
levels of the RuO$_6$ cluster broadened by a small 
intercluster hopping.
Then the highest occupied bands for the undoped compound
are the lower Hubbard bands of the three $t_{2g}$ character,
which are fully occupied.
Thus this material is a Mott insulator in the sense that
a charge gap is open between the upper and lower Hubbard bands.
When electrons are doped, the upper Hubbard bands are partially occupied,
which dominate the charge transport to give the negative Seebeck coefficient.
We expect that the doped electrons are easily localized partially
because of the  small intercluster hopping energy of 0.14 eV.
In addition, the electrons feel Hund's coupling to 
the three electrons in the lower Hubbard bands in hopping from one site
to another, which may further reduce the effective bandwidth and cause
the activation energy  in the mobility.
This doped Mott insulator is therefore difficult to be metallic, which
is a reason for the high resistivity and the Heikes-formula-type
Seebeck coefficient.

Finally we will make brief comments on the chemical properties of 
this double perovskite ruthenate.
(i) The chemical substitution for the $B$ and $B'$ sites can also
supply electrons to some extent. 
We made polycrystalline samples of 
Sr$_2$Er$_{1-x}$Ce$_x$RuO$_6$ and Sr$_2$ErRu$_{1-x}$Mo$_x$O$_6$,
and measured the resistivity and Seebeck coefficient.
We find that the high-temperature Seebeck coefficient is
roughly independent of temperature, and the magnitude is determined by 
the formal valence of the Ru ions.
The resistivity of Sr$_2$ErRu$_{1-x}$Mo$_x$O$_6$ is much higher than the
other serieses, indicating that the Ru-O network is a conduction path.
(ii) The title compound is highly stable in air up to 1000 K.
We fabricated a trial product of the thermoelectric module 
consisting of Ca$_3$Co$_4$O$_9$ and (Sr,La)$_2$ErRuO$_6$,
and examined the high-temperature stability.
The module is highly stable up to 1000 K both mechanically and
electronically, indicating that the title compound could be
a candidate for an n-type thermoelectric oxide if the 
resistivity could be reduced substantially.
(iii) We examined the substitution of 3d elements for Ru, and found that
Cu and Zn ions were partially substituted for Ru to decrease
resistivity. However, no trace of superconductivity was detected in our experiment.

\section{summary}
We  have prepared polycrystalline samples 
of (Sr$_{1-x}$La$_x$)$_2$ErRuO$_6$ and 
(Sr$_{1-x}$La$_x$)$_2$YRuO$_6$.
The x-ray absorption and susceptibility measurements
have clarified that  the La substitution for Sr
creates Ru$^{4+}$  in the low spin state,
and indicates that the conduction bands 
are the upper Hubbard $t_{2g}$ manifolds.
For $x$=0.1 at 800 K in air,
the Seebeck coefficient is negative and large 
($-$160 $\mu$V/K) , and 
the thermal conductivity shows a low value of 7 mW/cmK.
These two values are quite favorable as a thermoelectric material,
and are one of the best data among the thermoelectric oxides.
One last drawback is its high resistivity, which comes from a small 
transfer hopping between the RuO$_6$ clusters.
If the transfer hopping were improved significantly by properly
substituting the Sr or Y sites, 
the ordered ruthenates could be promising candidates 
for an n-type thermoelectric oxide.

We would like to thank T. Suzuki for collaboration at an early stage of
the Cu-substitution, and R. Funahashi and S. Urata for making a
thermoelectric power generator with the title compound.
We also appreciate M. Namba for giving us unpublished data
for the specific heat of (Sr$_{1-x}$La$_x$)$_2$YRuO$_6$, and A. Yamamoto
for fruitful discussion on the valence state of ruthenate.
This work was partially supported by the S-type Research Project, 
KEK Photon Factory (No. 2009S2-008), by the collaboration with NGK
Insulators Ltd., and by ALCA, Japan Science and Technology Agency.

\bibliographystyle{apsrev4-1}

%

\end{document}